\documentclass[onecolumn]{aa} 
\usepackage{graphicx}
\usepackage{sidecap}
\usepackage{txfonts}
\usepackage{natbib}
\bibpunct{(}{)}{;}{a}{}{,} 
\begin{document}
%
%

\title{Evolution and Nucleosynthesis of Extremely Metal Poor\\
  \& Metal-Free Low- and Intermediate-Mass Stars\\
  I: Stellar Yield Tables and the CEMPs\thanks{Tables 1 to 6 are only
    available in electronic form at the CDS via anonymous ftp to
    cdsarc.u-strasbg.fr (130.79.128.5) or via
    http://cdsweb.u-strasbg.fr/cgi-bin/qcat?J/A+A/}}

   \subtitle{}

   \author{S. W. Campbell
          \inst{1,} \inst{2}
          \and
          J. C. Lattanzio\inst{2}
          }

   \institute{Academia Sinica Institute of Astronomy and Astrophysics,
                 P.O. Box 23-141, Taipei 10617, Taiwan\\
              \email{simcam@asiaa.sinica.edu.tw}
         \and 
             Centre for Stellar and Planetary Astrophysics,
                 School of Mathematical Sciences,
                 Monash University,
                 Melbourne, Australia 3800\\
             \email{john.lattanzio@sci.monash.edu.au}
             }

   \date{Received 18 February 2008 / Accepted 23 August 2008}

 
  \abstract
  {The growing body of spectral observations of the extremely metal-poor
    (EMP) stars in the Galactic Halo provides constraints on theoretical
    studies of the chemical and stellar evolution of the early Universe.}
  {To calculate yields for EMP stars for use in chemical evolution
    calculations and to test whether such models can account for some of
    the recent abundance observations of EMP stars, in particular the
    highly C-rich EMP (CEMP) halo stars.}
  {We modify an existing 1D stellar structure code to include
    time-dependent mixing in a diffusion approximation. Using this code and
    a post-processing nucleosynthesis code we calculate the structural
    evolution and nucleosynthesis of a grid of models covering the
    metallicity range: $-6.5 \leq$ [Fe/H] $\leq -3.0$ (plus Z=0), and mass
    range: $0.85 \leq M \leq 3.0$ M$_\odot$, amounting to 20 stars in
    total. }
  { Many of the models experience violent nuclear burning episodes not seen
    at higher metallicities. We refer to these events as `Dual Flashes'
    since they are characterised by nearly simultaneous peaks in both
    hydrogen and helium burning. These events have been reported by
    previous studies. Some of the material processed by the Dual Flashes is
    dredged up causing significant surface pollution with a distinct
    chemical composition.  We have calculated the entire evolution of the
    Z=0 and EMP models, from the ZAMS to the end of the TPAGB, including
    extensive nucleosynthesis.  In this paper, the first of a series
    describing and analysing this large data set, we present the resulting
    stellar yields.  Although subject to many uncertainties these are, as
    far as we are aware, the only yields currently available in this mass
    and metallicity range. We also analyse the yields in terms of C and N,
    comparing them to the observed CEMP abundances. At the lowest
    metallicities ([Fe/H] $\lesssim -4.0$) we find the yields to contain
    $\sim 1$ to 2 dex too much carbon, in agreement with all previous
    studies. At higher metallicities ([Fe/H] $\sim -3.0$), where the
    observed data set is much larger, all our models produce yields with
    [C/Fe] values consistent with those observed in the most C-rich CEMPs.
    However it is only the low-mass models that undergo the Dual Shell
    Flash (which occurs at the start of the TPAGB) that can best reproduce
    the C \textit{and} N observations. Normal Third Dredge-Up can not
    reproduce the observations because at these metallicities intermediate
    mass models ($M \gtrsim 2$\,M$_\odot$) suffer HBB which converts the C
    to N thus lowering [C/N] well below the observations, whilst if TDU
    \textit{were} to occur in the low-mass ($M \leq 1$\,M$_\odot$) models
    (we do not find it to occur in our models), the yields would be
    expected to be C-rich only, which is at odds with the `dual pollution'
    of C and N generally observed in the CEMPs. Interestingly events
    similar to the EMP Dual Flashes have been proposed to explain objects
    similarly containing a dual pollution of C and N -- the `Blue Hook'
    stars and the `Born Again AGB' stars.  We also find that the proportion
    of CEMP stars should continue to increase at lower metallicities, based
    on the results that some of the low mass EMP models already have
    polluted surfaces by the HB phase, and that there are more C-producing
    evolutionary episodes at these metallicities.  Finally we note that
    there is a need for multidimensional fluid dynamics calculations of the
    Dual Flash events, to ascertain whether the overproduction of C and N
    at ultra-low metallicities found by all studies is an artifact of the
    1D treatment.}
  {}


   \titlerunning{Yields from Low-Mass EMP Stars}
   \maketitle
%
%
%
\section{Introduction}
%
In terms of chemical pollution the Extremely Metal Poor (EMP, [Fe/H]
$\lesssim -2.5$) Galactic Halo stars are the most ancient objects currently
known in the Universe. Observations show that their metallicities
reach as low as [Fe/H] $=-5.5$ (\citealt{Frebel2005}) -- far below the
metallicities measured in damped Ly$\alpha$ systems for example. Studying
the EMP stars is thus crucial to understanding the chemical evolution of
the early Universe. They provide the best link we have to the elusive first
generation of stars (Pop III).  Indeed, the surface abundances of
individual EMP stars are enriched to such a small degree that they may
reflect the composition of the ejecta of only a few (or even one) Pop III
supernovae. The chemical information from these stars should eventually
aid in our understanding of the First Stars, providing indirect evidence
for the chemical evolution of the Galaxy and, importantly, the first
stellar mass function (FMF), which is still very uncertain.  Deducing the
FMF is also important in terms of the Epoch of Reionisation.  For these
reasons this field is often referred to as `Near Field Cosmology'. The
nearness of the EMP stars (relative to ancient objects at high redshift)
also means that extremely detailed information can be collected from their
spectra using current instruments.

The discovery of Galactic Halo EMPs has naturally led to a renewed interest
in the theoretical modelling of Population III and low-metallicity stars.
In particular the subset of these ancient stars that is observed to contain
large amounts of carbon, the C-rich EMPs (CEMPs, which we define here as
stars with [C/Fe] $>+0.7$, also see Fig. \ref{CFEobs}) has attracted much
stellar modelling work since their abundance patterns are difficult to
explain with standard stellar evolution. These interesting objects also
appear to comprise a large proportion of the EMPs ($\sim 10 \rightarrow
20\%$; see eg.  \citealt{Beers2005, Cohen2005, Lucatello2006}), suggesting
that an additional (or modified) source of C production was active in the
early Universe. The CEMPs also display variation in a range of other
elements, such as s-process species (eg. \citealt{Goswami2006, Aoki2006,
  Sivarani2006}; also see \citealt{Beers2005} for an overview of the
observations). A number of theories have been proposed to explain the
various abundance patterns seen in CEMPs, ranging from pre-formation
pollution via Pop III supernovae (eg.  \citealt{Shigeyama1998,
  Limongi2003}) to self-pollution through peculiar evolutionary events (eg.
\citealt{Fujimoto2000, Weiss2004}) to binary mass transfer (eg.
\citealt{Suda2004}). 

In the current study we focus on the nature of the chemical pollution
produced by low- and intermediate-mass (LM and IM) EMP and Z=0 stars. As
these stars generally don't produce the heavier elements (except possibly
for s-process elements), we assume that they have formed from gas clouds
already enriched by Pop III supernovae that produced the near scaled-solar
abundances of many elements seen in the `normal' (non C-rich) EMP Halos
stars. In terms of producing the excessive light element pollution evident
in the CEMPs we keep an open mind as to whether they may have received
their compositions through self-pollution events or through binary
mass-transfer (be it from wind accretion or Roche lobe overflow).

Stellar modelling of low-mass EMP stars started in the early 1960s, when
\cite{Ezer1961} calculated ZAMS models of Z=0 stars. In the early 1980s it
was realised that low-mass EMP stars may suffer H-ingestion during the core
He flash at the tip of the Red Giant Branch (\citealt{Dantona1982}).  About
ten years later this was confirmed by detailed simulations
(\citealt{Fujimoto1990, Hollowell1990}). The H-ingestion during these
evolutionary events -- which are peculiar to models of low-mass Z=0 and EMP
stars -- results in a rapid burning of the protons, a H-flash. More recent
work has investigated the dependence of the H-ingestion flashes on physical
inputs (eg.  \citealt{Schlattl2001}), compared the chemical signatures from
the simulations with observations of CEMPs (eg. \citealt{Fujimoto2000,
  Chieffi2001}) and explored the possibility of initiating the s-process at
zero metallicity (\citealt{Goriely2001}). 

Although some grids of EMP and Z=0 models have already been computed the
studies have either ignored the H-ingestion events (eg.
\citealt{Cassisi1993, Marigo2001}) and/or have not calculated all of the
AGB evolution. In the current study we aim to provide a homogeneous set of
yields that include the nucleosynthetic effects of the EMP H-ingestion
events (which we refer to below as `Dual Flashes'), AGB Third Dredge-Up
(TDU, the periodic mixing up of He-burning products after a thermal pulse)
and AGB Hot Bottom Burning (HBB, the H-burning of the convective envelope).
To this end we have undertaken a broad exploration of EMP ($-6.5 \leq$
[Fe/H] $\leq -3.0$) and Z=0 stellar evolution and nucleosynthesis in the
low and intermediate mass regime ($0.85\leq M \leq 3.0$ M$_\odot$). We
include (for the first time) evolutionary and nucleosynthesis calculations
from ZAMS to the end of the thermally-pulsing AGB phase (TPAGB) involving
74 species, as well as chemical yield tables.  With this homogeneous set of
models we hope to shed some light on whether or not 1D stellar models can
help to explain some of the EMP halo star observations, and in particular
the CEMP abundance patterns.
\begin{figure}
\begin{center}
\resizebox{0.80\hsize}{!}{\includegraphics{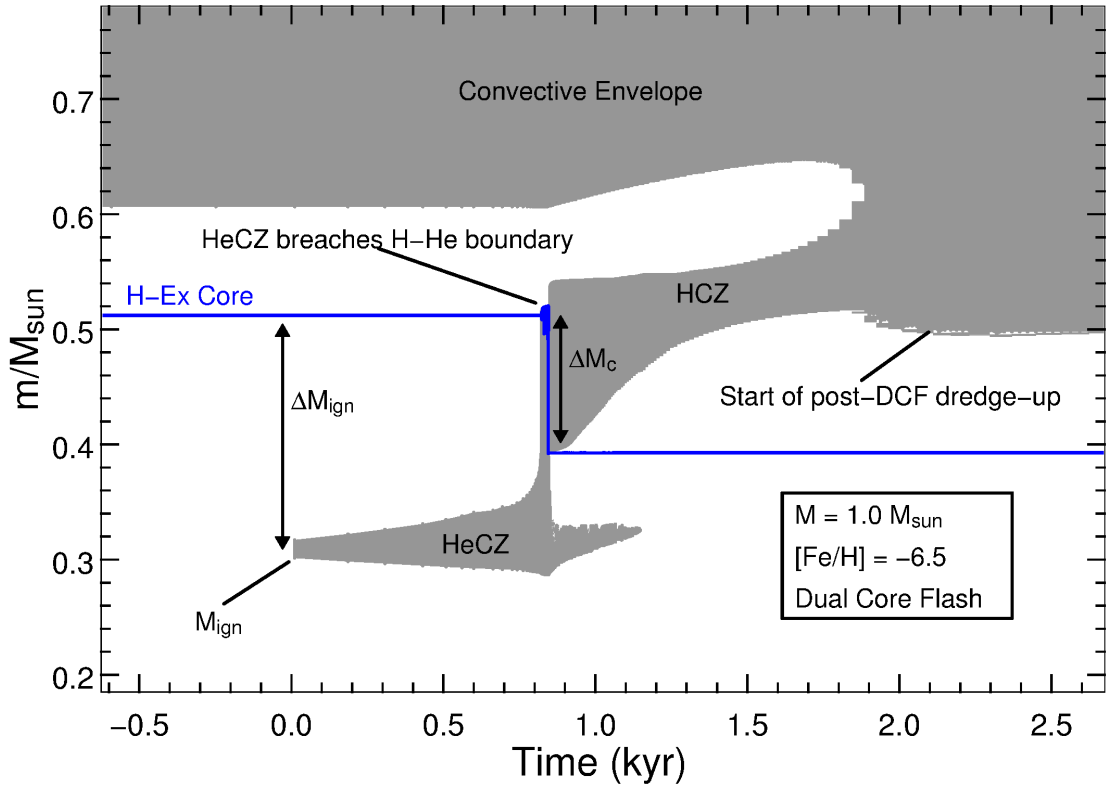}}
\caption{Example of one of the Dual Core Flash events. Convective regions are shown by
  grey shading whilst the mass location of the edge of the H-exhausted core
  is shown by the solid line (blue).}
\label{DCF}
\end{center}
\end{figure}
%
\section{Method}
\label{method}
Our simulations were performed utilising two numerical codes -- a
stellar structure code and post-processing nucleosynthesis code.

The stellar structure code used was the Monash version of the Monash/Mount
Stromlo stellar evolution code (MONSTAR, see eg.  \citealt{Wood1981,
  Frost1996}). The code is largely a standard 1D code that utilises the
Henyey-matrix method (a modified Newton-Raphson method) for solving the
stellar structure equations. For the present study the instantaneous
convective mixing routine was replaced by a time-dependent (diffusive)
mixing routine (similar to that described by \citealt{Meynet2004}). This
change was necessary due to the violent evolutionary events (the H-flashes)
that occur in models of Z=0 and EMP stars, where the timescale for mixing
becomes comparable to the evolutionary time steps. Our numerical scheme
remained the same, `partially simultaneous', which means mixing and
structure are calculated within each iteration (see
\citealt{Stancliffe2006} for a comparison of different methods).  

Opacities have been updated to those from \cite{Iglesias1996} (for
mid-range temperatures) and \cite{Ferguson2005} (for low temperatures).
Convective boundaries were always defined by the Schwarzschild criterion --
ie. the search for a neutral convective boundary (see \citealt{Frost1996})
was not performed and no overshoot was applied. 

A key problem with modelling EMP stars is the unknown driver(s) of mass
loss.  We have used the empirical mass-loss formula of \cite{Reimers1975}
during the RGB.  We believe this to be acceptable because the mass lost
prior to the AGB is very small due to the short-lived giant branches at
these metallicities (see eg.  \citealt{Dantona1982}). For the AGB we use
the formula of \cite{Vassiliadis1993}. As described below, all the models
experience some self-pollution -- and always before or at the very
beginning of the TPAGB phase. We find that the surfaces of the AGB models
usually have metallicities approaching that of the LMC or even Solar (as
defined by $Z = 1-X-Y$ rather than Fe -- they are still Fe-poor). Thus,
since the stellar surfaces have (some of) the ingredients needed to form
grains, we argue that using a standard mass loss formula is warranted, at
least as a first approximation. We note that metallicity is also indirectly
taken into account by the mass loss formulae, since they depend on bulk
stellar properties (such as radius, luminosity, pulsation period), which
vary significantly with metallicity.

The nucleosynthesis calculations were made with the Monash Stellar
Nucleosynthesis code (MONSOON), a post-processing code which takes input
from the MONSTAR code (eg. density, temperature, convective velocities). It
solves a network of 506 nuclear reactions involving 74 nuclear species (see
eg. \citealt{Cannon1993, Lattanzio1996, Lugaro2004}). The yield tables
contain a reduced number of species because we have excluded isotopes with
very short lifetimes (and thus have negligible yields).

Initial composition for the Z=0 models was taken from the Standard Big Bang
nucleosynthesis calculations of \cite{Coc2004}, whilst the initial
composition for the EMP models was derived by mixing the ejecta from a
20\,M$_\odot$ Z=0 supernova calculation (Limongi 2002, private
communication) with varying amounts of Big Bang material from
\cite{Coc2004} to reach the desired [Fe/H] values (for example
$10^6$\,M$_\odot$ of Big Bang material was required for [Fe/H] $=-4.0$).
The main difference between this composition and that of a scaled-solar
composition is an underabundance of N, since the supernova calculation did
not produce much of this element.  This was found to have little effect
however since whenever the CNO cycle operates it quickly converts much
of the C to N anyway ([C/Fe] is $\sim 0$ initially).

Our grid of models covers the mass range: $M = 0.85,1.0,2.0,3.0$\,M$_\odot$
and the metallicity range: [Fe/H] $=-6.5,-5.45,-4.0,-3.0$, plus Z=0.
\begin{figure}
\begin{center}
\resizebox{0.80\hsize}{!}{\includegraphics{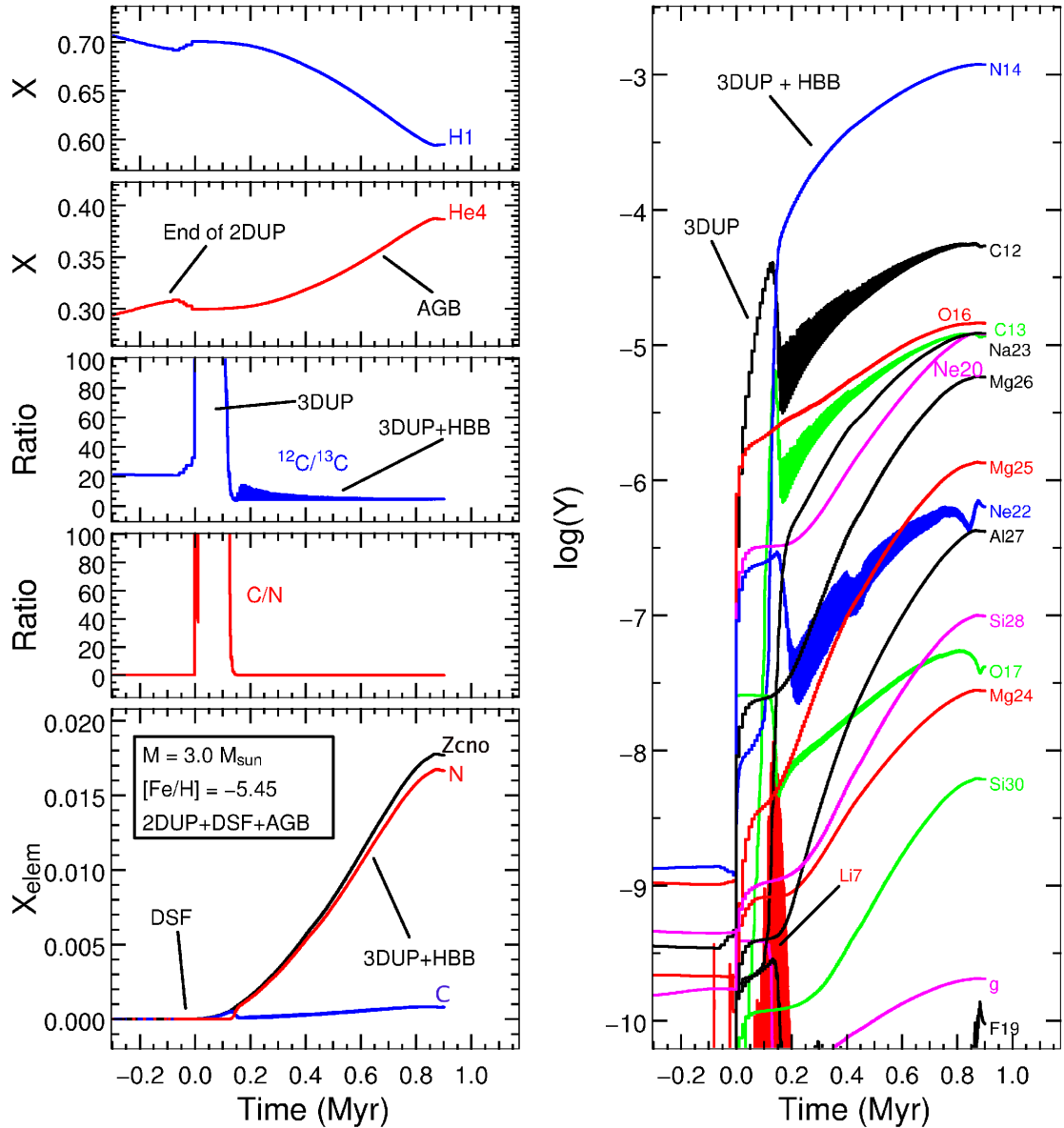}}
\caption{Time evolution of the surface composition in the [Fe/H] $=-5.45$,
  3\,M$_\odot$ model, for selected species. This model is a member of our
  self-pollution Group 3 (see text and Fig. \ref{HRDs} for details).  The
  rich nucleosynthesis arising from TDU and HBB is seen in the right-hand
  panel.  This chemical signature by far dominates that of the DSF in this
  case (see bottom left panel).}
\label{SRF}
\end{center}
\end{figure}
\section{Results and Discussion}
\subsection{The Dual Flash Events}
\label{DF}
%
As mentioned above it has long been known that theoretical models of Z=0
stars (and EMP stars) undergo violent evolutionary episodes not seen at
higher metallicities.  

The most severe of these evolutionary events occurs during the core He
flash of low-mass stars (with [Fe/H] $\lesssim -4.5$, see
\citealt{Fujimoto2000}). In this event the normal flash-driven convection
zone breaks out of the He-rich core. Thus H-rich material is mixed down to
regions of high temperature, producing a secondary flash: a H-flash. This
flash reaches luminosities comparable to the core He flash itself and
exists concurrently with the He flash. Thus we refer to the combination of
these events as a `Dual Core Flash' (DCF). These events have been given
rather cumbersome names in the literature to date: eg. Helium Flash Induced
Mixing (HEFM, \citealt{Schlattl2002}) and Helium Flash-Driven Deep Mixing
(He-FDDM, \citealt{Suda2004}). We propose the simpler name DCF to
illustrate the essential nature of the phenomenon. 

A similar event occurs in stellar models of higher mass and higher
(although still very low) metallicities.  In these cases it is the normal
flash-driven convection zone present during the first pulse (or first few
pulses) of the TPAGB phase that breach the H-He discontinuities. Again a
H-flash results, concomitant with the He shell flash, so we refer to this
event as a `Dual \emph{Shell} Flash' (DSF). \cite{Cassisi1996} appear to be
the first to have reported the occurrence of a DSF, although they were
unable to follow the evolution of the event. DSFs have since been reported
and modelled by a number of groups (eg. \citealt{Chieffi2001, Siess2002,
  Iwamoto2004}).

Both the DCF and DSF events are driven by the same phenomenon: ingestion of
protons into a hot region caused by expansions of He convective zones into
H-rich regions.  Thus our proposed nomenclature unifies them as being
``Dual Flashes'' and then distinguishes them again by referring to the
driving event: be it the core or shell flash. Furthermore, the term only
applies to H-flash events occurring at low metallicities.

Both the DCF and DSF have consequences for the surface composition of the
star since, in both cases, the convective envelope subsequently deepens and
mixes up the (processed) material overlying the H-burning shell.  In Fig.
\ref{DCF} we display an example DCF event (our 1\,M$_\odot$, [Fe/H] $=
-6.5$ model). Here we can see the normal Helium-Flash Convective
Zone (HeCZ) rapidly expanding into the H-rich regions above. The result is
that a second convective zone, which is H-rich (HCZ) is set up by the rapid
proton burning at its base (the H-flash). The HeCZ reduces in mass
due to a reduction in He burning and, in this case, the two convective zones
remain separated. A few thousand years later the convective envelope moves
in and dredges up most of the processed material located above the
H-exhausted core (labelled H-Ex Core in the figure; only the beginning of
this dredge-up is shown), causing a strong pollution of the stellar
envelope.  Since this model experiences no Third Dredge-Up during the AGB
phase the DCF dredge-up is the main contributor to the chemical yield.
 
Interestingly the Dual Flash phenomenon (and subsequent dredge-up) is
similar to the ``Late Hot Flasher'' scenario proposed to explain Blue Hook
stars in Galactic globular clusters, where white dwarves produced from RGB
stars with excessive mass loss experience a late core He-flash that engulfs
the remaining H-rich envelope (see eg. \citealt{Sweigart1997, Brown2001,
  Cassisi2003}). Another similar phenomenon is the ``Born Again'' scenario
for producing H-deficient post-AGB (PAGB) stars (such as Sakurai's Object;
see eg.  \citealt{Asplund1997}). In this case a star that has recently left
the AGB experiences a very late helium-shell flash that also engulfs the
remaining H-rich envelope, again producing a H-flash (see eg.
\citealt{Iben1983, Renzini1990, Herwig1999}). Both these scenarios lead to
surface pollution similar to that produced by the DF events since the
matter mixed up in these cases has also undergone H and He burning. We note
however that the envelope mass overlying the He convective zones is much
lower in both these cases ($\sim10^{-4}$ M$_\odot$) than in the EMP/Z=0
Dual Flash case ($\sim 10^{-1}$ M$_\odot$ or higher).

We discuss the chemical composition of the surface pollution resulting from
the Dual Flash events when we compare the C and N yields with observations
below (\S\ref{compareobs}).
\begin{table*}
  \caption{Part of the yield table for the Z=0 models -- selected species only. Initial composition 
    is included in the third column.}
\label{table-yieldsZERO}
\centering   
\begin{tabular}{ccccccc}
\hline \hline       
Nuclide&
A&
Initial&
0.85 M$_{\odot}$& 
1.0  M$_{\odot}$& 
2.0  M$_{\odot}$&
3.0  M$_{\odot}$\\
\hline
$^{1}$H&	1&	7.548E-01&	7.014E-01&	6.597E-01&	6.596E-01&	5.807E-01\\
$^{4}$He&	4&	2.450E-01&	2.976E-01&	3.295E-01&	3.367E-01&	4.066E-01\\
$^{7}$Li&	7&	3.130E-10&	4.704E-10&	1.263E-09&	4.491E-10&	4.887E-11\\
$^{12}$C&	12&	0.000E+00&	2.598E-05&	1.844E-03&	1.309E-04&	4.882E-04\\
$^{13}$C&	13&	0.000E+00&	7.778E-06&	3.619E-04&	3.034E-05&	1.152E-04\\
$^{14}$N&	14&	0.000E+00&	2.437E-04&	3.919E-03&	3.432E-03&	1.166E-02\\
$^{16}$O&	16&	0.000E+00&	5.034E-04&	4.333E-03&	4.885E-05&	1.516E-04\\
$^{19}$F&	19&	0.000E+00&	1.848E-09&	6.225E-06&	2.879E-10&	1.188E-09\\
$^{20}$Ne&	20&	0.000E+00&	2.485E-07&	1.726E-06&	2.737E-05&	1.386E-04\\
$^{23}$Na&	23&	0.000E+00&	1.291E-09&	1.131E-05&	1.294E-05&	9.539E-05\\
$^{24}$Mg&	24&	0.000E+00&	3.838E-11&	1.362E-06&	1.865E-07&	2.630E-07\\
$^{25}$Mg&	25&	0.000E+00&	1.459E-08&	3.166E-07&	1.756E-06&	1.562E-05\\
$^{26}$Mg&	26&	0.000E+00&	2.475E-08&	4.065E-08&	8.159E-06&	6.889E-05\\
$^{26}$Al&	26&	0.000E+00&	3.182E-11&	3.364E-10&	3.085E-07&	1.487E-06\\
$^{28}$Si&	28&	0.000E+00&	1.002E-07&	1.649E-11&	6.170E-07&	1.315E-06\\
$^{31}$P&	31&	0.000E+00&	2.200E-08&	2.071E-12&	5.219E-07&	1.117E-06\\
$^{32}$32&	32&	0.000E+00&	4.381E-09&	1.224E-12&	1.870E-07&	2.665E-07\\
\hline   
\end{tabular}
\end{table*}
%
%
\subsection{Nucleosynthetic Yields and Their Categorisation}
\begin{figure}
\centering
\resizebox{0.4\hsize}{!}{\includegraphics{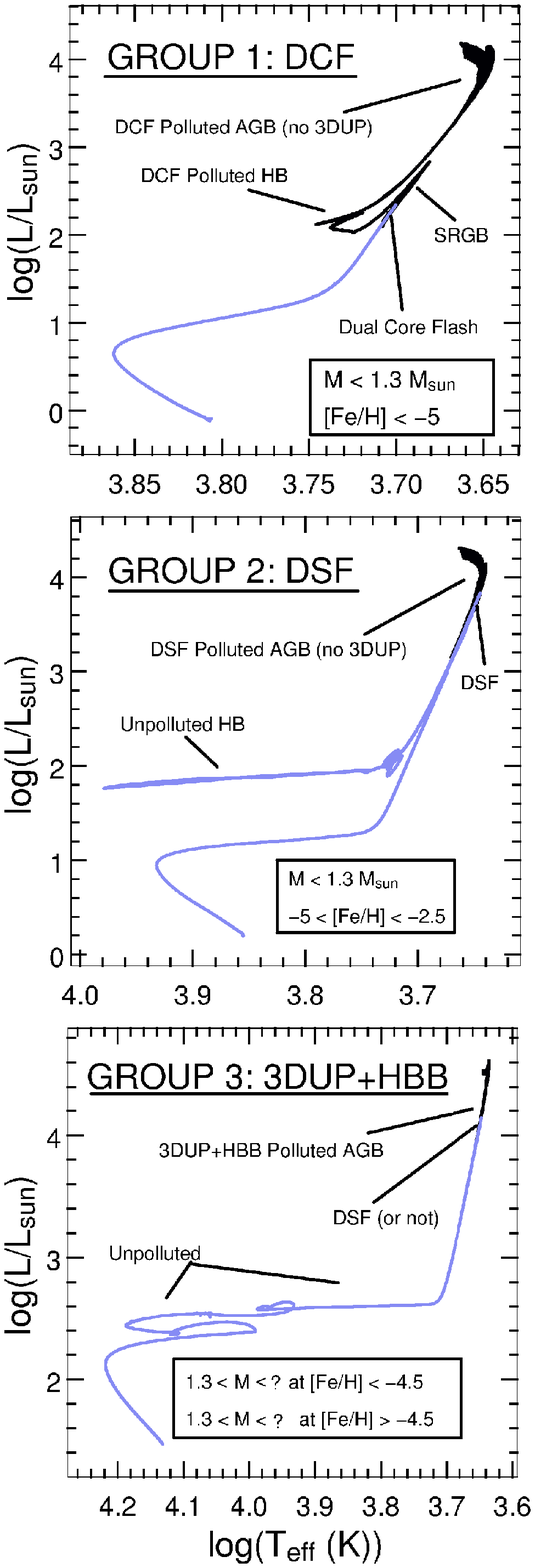}}
\caption{Displayed in each HR diagram is a representative example from our
  grid of models for each `Self-Pollution Group' (see text for details on
  the groups). Black lines indicate phases of the evolution in which the
  surface is strongly polluted with CNO nuclides (from the DCF, DSF or 3DU
  events).  Grey (light blue) lines indicate that the surface still retains
  the initial metal-poor composition. Evolutionary stages and
  self-pollution sources are marked, as are the mass and metallicity ranges
  of each group.  Question marks indicate unknown upper boundaries (due to
  the limited mass range of the current study).}
\label{HRDs}
\end{figure}
In Table \ref{table-yieldsZERO} we present a sample of the yields for the
models. A total of five tables with the same format as these are available
in their entirety online at the CDS (Tables 1 to 5; via anonymous ftp to
cdsarc.u-strasbg.fr (130.79.128.5) or via
http://cdsweb.u-strasbg.fr/cgi-bin/qcat?J/A+A/ ). Each table contains
yields for models with initial stellar masses of 0.85, 1.0, 2.0 and
3.0\,M$_\odot$ at each of the metallicities: [Fe/H]
$=-6.5,-5.45,-4.0,-3.0$, and Z=0. We list the yields for all the stable
species used in the nucleosynthesis calculations, which range from $^{1}$H
to $^{34}$S plus a small iron group (Ni, Co, Fe isotopes). We also provide
yields for the important radionuclides $^{26}$Al (which decays to
$^{26}$Mg) and $^{60}$Fe (which decays to $^{60}$Ni).  Yields are given in
mass fraction of each species in the total ejecta.  In the tables we also
give the initial compositions for each metallicity. The final masses
(remnant core masses) are given in Table 6 (also at CDS). Using this
information it is easy to convert the yields to any format. The yields are
calculated by integrating the mass of each species lost by the star over
its lifetime ($\tau_{*}$):
\begin{equation}
M_{i}^{tot}=\int_{0}^{\tau_{*}}X_{i}(t)\frac{dM}{dt}dt\label{eqn-YieldCalc}
\end{equation}
where $X_{i}$ is the mass fraction of species $i$. The total mass
of each species lost to the ISM, $M_{i}^{tot}$, is then scaled with
$M_{ej}$, the total mass lost by the star, to give the mass fractions.

In some cases our models failed to converge towards the end of the
AGB. This is a common problem with stellar codes. Often there was
very little mass left in the envelope so this was just added to the
yields. However in some cases there was enough mass left that it would
not have been lost in one interpulse period. In these cases we performed
a short synthetic evolution calculation for the remaining thermal pulses
(including third dredge-up and core growth) to complete the evolution,
following the method of \cite{Karakas2003}. Yields were then calculated
taking into account this extra mass loss. In most models the number of thermal
pulses calculated in this way was $\lesssim 8$. This represents between
$\sim 1$ and $10\%$ of the total number of thermal pulses in most
of our models. Thus the synthetic pulses generally have a minor impact
on the yields.

As an example of the nucleosynthesis we show in Fig. \ref{SRF} the
results for our $M=3$ M$_\odot$ and [Fe/H] $=-5.45$ model. Here it is
TDU and Hot Bottom Burning that are the main contributors to the yield of
the star. Indeed, the chemical signature arising from the DSF occurring at
the start of the AGB is totally erased by these normal AGB evolutionary
episodes. In particular the CN cycling product $^{14}$N dominates the
surface composition during most of the AGB (in terms of metallic species).
The $^{12}$C$/^{13}$C and C/N ratios quickly approach equilibrium values
once the (strong) HBB starts.

We summarise the self-pollution episodes over the whole grid of models by
dividing them into three categories, defined by the evolutionary
events/phases that dominate the chemical signature in the yields:
\begin{itemize}
\item{Group 1 yields are dominated by the DCF events}
\item{Group 2 are dominated by DSF events}
\item{Group 3 are dominated by TDU+HBB}
\end{itemize}
Members of the DCF group have polluted surfaces during the horizontal
branch phase onwards whilst the members of the DSF group have polluted
surfaces after the first few pulses of the AGB (see Fig. \ref{HRDs}). In
the low mass models ($M \leq 1.0$ M$_\odot$) this pollution occurs despite
the lack of TDU.  Thus our models predict a greater proportion of C-rich
stars at extremely low metallicity, since these Dual Flash events do not
occur at higher metallicities. 

\begin{figure}
\begin{center}
\resizebox{0.85\hsize}{!}{\includegraphics{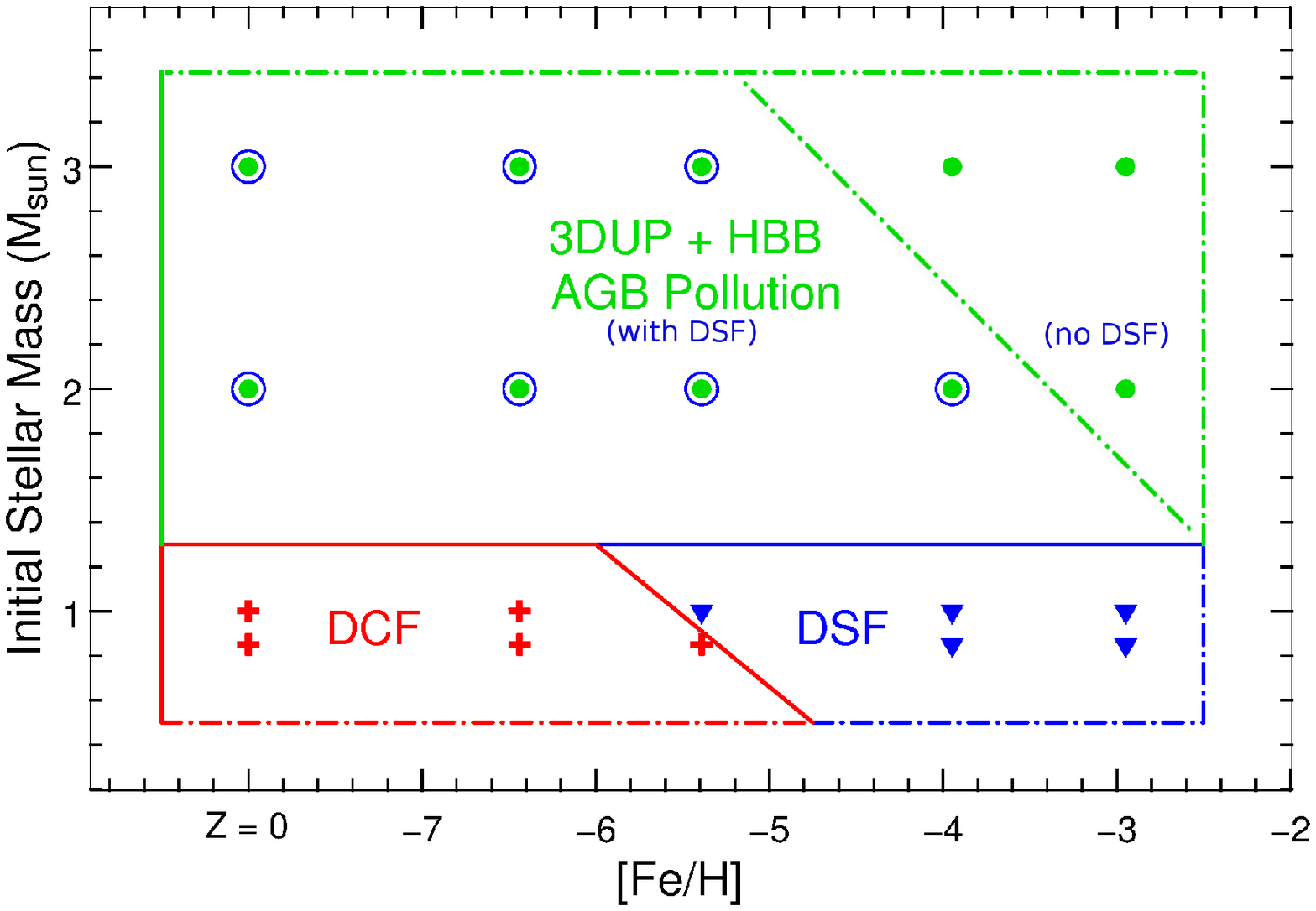}}
\caption{Mass-metallicity diagram summarising the dominant sources of
  pollution in the yields. Each symbol represents a single stellar model.
  Crosses (red) represent the DCF self-pollution Group 1, filled triangles
  (blue) the DSF Group 2 and filled circles (green) the AGB pollution Group
  3. The open circles (blue) around the filled circles (green) indicate
  intermediate mass models that experienced DSFs. Pollution from TDU (and
  HBB) easily dominates the pollution from the DSF events at IM mass so the
  yields of these models fall into the AGB group. The Z=0 models are
  included at [Fe/H] $= -8$.}
\label{MMD}
\end{center}
\end{figure}

In Fig. \ref{MMD} we show the self-pollution groups in a mass-metallicity
diagram. Here it is clear that the yields of all the intermediate mass
models ($M \geq 2.0$ M$_\odot$) that suffer surface pollution resulting
from Dual Shell Flashes are actually dominated by the pollution occurring
during the AGB phase. In other words the AGB pollution erases the DSF
pollution. It can also be seen that the DCF events are limited to ultra
metal poor low-mass models, whilst the DSF events occur at higher
metallicities (for low mass models). We note that \cite{Fujimoto2000} also
provide a mass-metallicity diagram for their low-metallicity study (their
Fig. 2).  Our diagram is qualitatively similar to theirs.  One point of
difference is that our boundary for the AGB-DSF models (filled circles with
open circles around them in Fig. \ref{MMD}) is at a lower metallicity.
This may be due to the fact that we adopt a `hard' Schwarzschild convective
boundary in our models, although we are unsure if \citealt{Fujimoto2000}
used any overshoot or not. A further small difference is that our diagram
shows a mass dependence in addition to the metallicity dependency for the
pollution event boundaries (diagonal lines in Fig. \ref{MMD}).
\begin{figure}
\begin{center}
\resizebox{0.85\hsize}{!}{\includegraphics{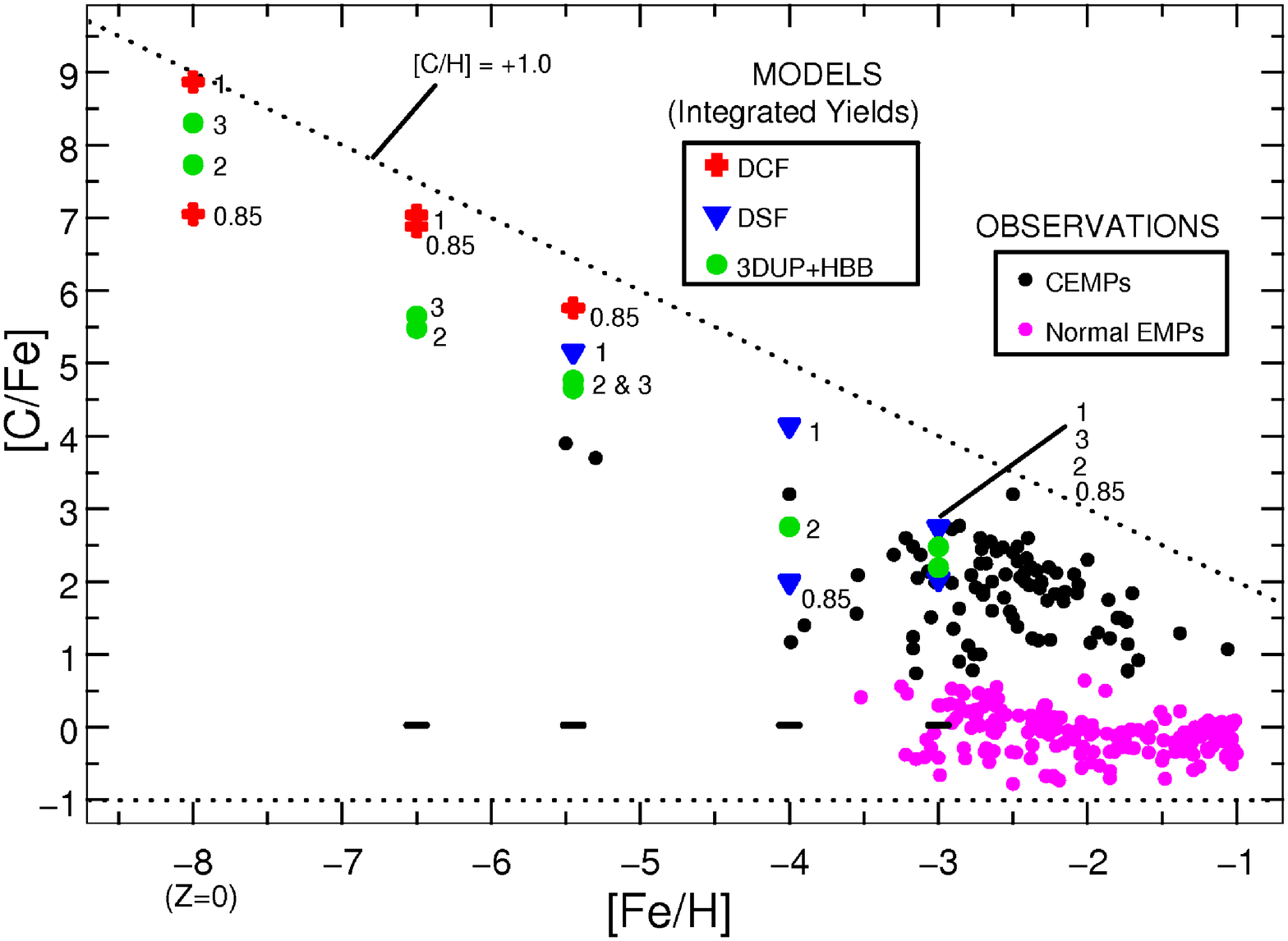}}
\caption{Carbon yields from our models and observations of EMP stars.
  Stars with [C/Fe] $> +0.7$ are shown in black (the CEMPs), those with
  [C/Fe] $< +0.7$ are shown in grey (magenta).  The short horizontal lines
  indicate the starting composition of the models (all have [C/Fe] $\sim
  0$, except for the Z=0 models). We have plotted the Z=0 model yields at
  [Fe/H] $= -8$.  The two most metal-poor stars can be seen at [Fe/H] $\sim
  -5.5$ (both are C-rich). An upper envelope to the self-pollution of the
  models -- and the observations -- is marked by the dotted line at [C/H]
  $= +1.0$.  The yields from our models are colour- and shape-coded to
  highlight the different episodes that produced the bulk of the pollution
  in each yield (see text for details). Numbers beside each yield marker
  indicate initial stellar mass, in M$_\odot$. The 3\,M$_\odot$ model at
  this [Fe/H] $=-4.0$ is missing due to a data loss. Observational data are
  from \cite{Frebel2006, Spite2006, Aoki2007, Beers2007, Cohen2006,
    Christlieb2004} and \cite{Aoki2006a}.}
\label{CFEobs}
\end{center}
\end{figure}
%
\subsection{Preliminary Comparisons with Observations: C \& N}
\label{compareobs}
%
In Fig. \ref{CFEobs} we compare the carbon yields from our entire grid of
models with the observed [C/Fe] abundances in EMP halo stars. It can be
seen that the yields are universally C-rich. Thus there is a qualitative
agreement between the models and the observations in terms of C, as found
by previous studies. Another interesting feature of this diagram is that
the model yields predict [C/Fe] to continue increasing towards lower and
lower metallicities (stars of such low metallicity may show up in future
surveys). Furthermore, taking into account the evolutionary stage at which
the surface pollution is gained in the lower mass models ($M = 0.85$ and
1.0 M$_\odot$) -- ie. the DCF events rather than the AGB -- the models also
predict a higher proportion of C-rich stars at lower and lower
metallicities. This is due to the fact that these stars already have
self-polluted surfaces during the HB stage -- which has a lifetime roughly
1 order of magnitude longer than the AGB phase.

Previous studies have also attempted to quantitatively reproduce the
abundances observed in some CEMP stars, in particular C and N. However they
have universally found that the 1D models produce by far too much of these
elements. The early study by \cite{Hollowell1990} reported their resultant
post-DCF surface abundance of nitrogen to be $\sim 10^2$ times that
observed in the EMP star \object{CD$-$38 245}. \cite{Schlattl2002} find about
the same two orders of magnitude overproduction of both N and C, whilst
\cite{Iwamoto2004} find their [Fe/H] $= -2.7$ models produce $\sim 1$ to 2
dex too much C and N than observed. 

\begin{figure}
\begin{center}
\resizebox{0.85\hsize}{!}{\includegraphics{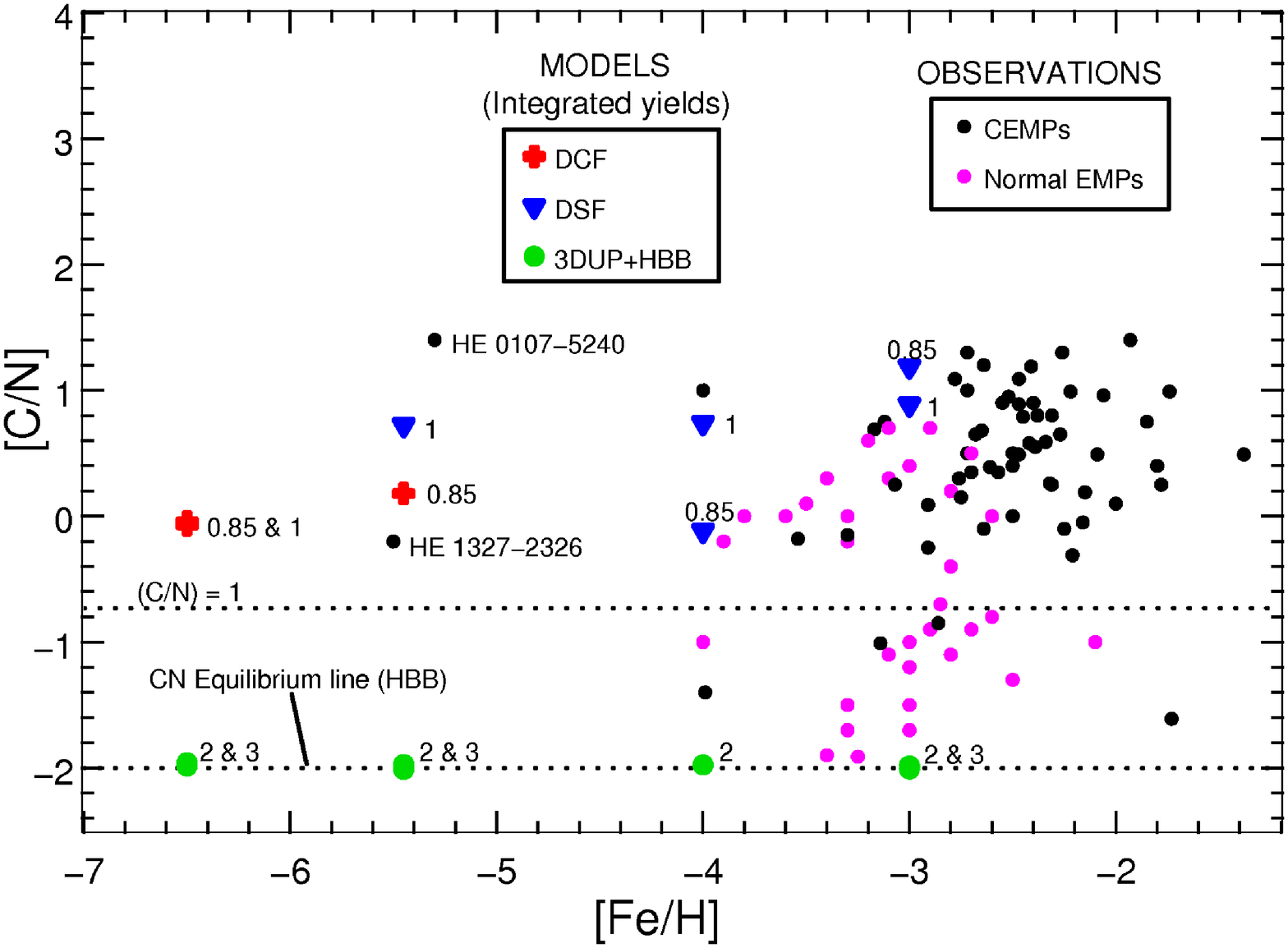}}
\caption{Comparing the [C/N] ratios in the yields of our models with those
  of the observations. See Fig. \ref{CFEobs} for the observational data
  sources and the definitions of CEMPs and `normal' EMPs. We have marked in
  the CN equilibrium line (for HBB) at [C/N] $= -2$ (ie.  C/N $= 0.05$).
  All the models that experienced HBB fall on this line. We have also
  marked in a dotted line at [C/N] $\sim -0.8$ which is where C/N $\sim 1$,
  so that C dominates N above this line. For reference [C/N] $= +1$ is
  equivalent to C/N $\sim 50$. Note that we have not plotted the Z=0 models
  in this case.}
\label{CN}
\end{center}
\end{figure}

In Fig. \ref{CFEobs} it can be seen that our models produce similar
overabundances at the lowest metallicities.  For instance, at [Fe/H] $=
-5.5$, the current lower limit of the observations, our models produce
$\sim 1$ to 2 dex too much C. Our findings thus concur with the previous
studies -- the DCF self-pollution scenario is not viable for these stars
(also see \citealt{Picardi2004}), at least in terms of current 1D models.
We note however that the overabundance problem may be alleviated somewhat
in a binary mass-transfer scenario whereby the material would be diluted in
the secondary star's envelope. Interestingly at this metallicity it is the
higher mass models (2 and 3\,M$_\odot$), which have yields dominated by AGB
products, that are the closest to the observations -- their yields are only
$\sim 0.7$ dex overabundant in C.  However these models have undergone
strong HBB (see Fig. \ref{SRF}), which has made their yields highly
N-rich and thus pushed their [C/N] ratios well below the CEMP observations
(Fig. \ref{CN}).

At [Fe/H] $= -4.0$ our yields are more consistent with the
observations, although the 1\,M$_\odot$ model still produces $\sim 1$ dex
too much C. It is important to note here that the observational data
samples at these ultra low metallicities are still very small (eg. there
are only two stars at [Fe/H] $\sim -5.5$). 

At [Fe/H] $= -3.0$ the data sample is much larger.  Here we find that our
carbon results are much more in line with observations. At this metallicity
DCFs are not found to occur (see Figs. \ref{HRDs} and \ref{MMD}). Instead
the lower mass models experience Dual \textit{Shell} Flashes at the start
of the AGB. Since they do not show TDU the abundance patterns in their
yields are primarily from the DSFs. The higher mass models do not undergo
any Dual Flashes at all, they experience normal AGB TDU and strong HBB
(even at 2\,M$_\odot$).  However the low [C/N] ratios in the yields from
these IM models are again too low compared to the observations.  Thus,
based on the two constraints of [C/Fe] and [C/N], it is our low mass models
which undergo DSFs that provide the best fit to the observations. We note
that our results are in contrast with the models of \cite{Iwamoto2004}, who
found their [Fe/H] $= -2.7$ models to produce $\sim 1$ to 2 dex too much C
and N.

We display some [C/N] ratios from EMP Galactic Halo star observations along
with our yield abundances in Fig. \ref{CN}. An interesting feature of the
observations is that the bulk of the CEMPs have [C/N] values ranging from
Solar to $\sim 1$ dex super-solar (which corresponds to C/N $\sim 50$). In
contrast the `normal' (non C-rich) EMPs show a very large spread, from 2
dex sub-solar to 1 dex super-solar. An important point here is that the
CEMPs are simultaneously C- and N-rich -- although they generally have
lower N abundances than C, they are also strongly enriched in N. The newly
constructed `Stellar Abundances for Galactic Archeology' (SAGA) Internet
database of \cite{Suda2008}, which contains more observational data, also
shows these features. 

The `dual pollution' of C and N is qualitatively similar to the pollution
found to arise from the Dual Flashes, a result that has been reported by a
number of previous studies (eg.  \citealt{Hollowell1990, Schlattl2002}).
Significantly this contrasts with the chemical pollution expected from TDU
in low mass stars, which would be predominately C-rich. As mentioned above,
the yields from the models which suffer HBB are heavily enriched in N -- so
much so that N rather than C dominates their composition -- in contrast to
the DF-polluted yields. Thus the yields from intermediate mass AGB stars
can not reproduce the bulk of the CEMP observations (we note that there are
a few CEMPs located below the C/N = 1 line; also see \citealt{Johnson2007}
for an analysis of the apparent paucity of N-rich CEMPs). 

In Fig. \ref{CN} it can also be seen that the spread of [C/N] values in the
yields dominated by the DF events (ie. the low-mass models) are reasonably
consistent with the spread of the CEMP observations.  Again at the lowest
metallicities ([Fe/H] $= -5.5, -4.0$) we are dealing with very small
samples (2 or 3 stars). The main discrepancy is \object{HE 0107$-$5240}
which has a [C/N] ratio significantly above that of our models of
comparable metallicity. We note that \cite{Venn2008} have recently
suggested that some of the most metal-deficient Halo stars, such as this
one, may actually be low-metallicity analogues of the `chemically peculiar'
stars (which show dust depletion of particular elements, such as Fe) and
thus may be intrinsically more metal-rich than currently thought.

The yields from our higher metallicity low-mass models ([Fe/H] $= -3.0$)
are at the high end of the observations ([C/N] $\sim 1$).  Looking back at
Fig. \ref{CFEobs} the reason for this seems to be that our models produce C
at the upper end of the CEMP distribution. One possible solution to this is
dilution of the C-rich material with scaled-solar C/Fe in a binary
mass-transfer scenario.  Another possibility is that the Dual Shell Flash
events -- which we stress are the only events that come close to
simultaneously reproducing the [C/Fe] and [C/N] observations -- may in
reality produce a distribution of C (and N) pollution that is not reflected
in the models.

It is interesting to note that other objects also show surfaces
simultaneously enriched in C and N. Recent observations of potential `Late
Hot Flasher' stars (see \S\ref{DF}) show this feature (eg.
\citealt{Lanz2004}), as does Sakurai's Object, one of the few known
`Born-Again' AGB stars (\citealt{Asplund1999}). The fact that the chemical
patterns in these two types of objects are thought to arise from He-flash
induced H-ingestion episodes lends weight to the suggestion that the CEMP
abundance patterns are a product of the similar phenomenon of EMP Dual
Flashes.

Although the Dual Flash [C/Fe] and [C/N] abundances in the yields compare
reasonably with the observations, the lifetimes of these stars also need to
be taken into account.  We find models with initial masses of 0.85
M$_\odot$ (ie.  the low-mass edge of our grid) have lifetimes of $\sim 10$
Gyr. This is comparable to the age of the Galactic Halo Globular clusters
so these stars may still be `living' today. Conversely the 1, 2 and 3
M$_\odot$ models, which have main sequence lifetimes of $\sim 5.7$, $\sim
0.6$ and $\sim 0.2$ Gyr respectively, would not have lasted to the present
day. They could however have contributed to the chemical enrichment of the
CEMP population through binary mass transfers (see eg.
\citealt{Suda2004, Lucatello2005, Stancliffe2007, Lugaro2008}).

Finally we note that here we have only discussed the chemical signatures of
C and N. We shall provide further comparisons and analysis of other
elements and abundance patterns in future papers in this series.
\subsection{Uncertainties \& the Way Forward}

The results presented here naturally contain many uncertainties, especially
since they are the first attempt at detailed yields in this mass and
metallicity range, and particularly because EMP stellar evolution is so
challenging. 

Some sources of uncertainties include: the unknown mass-loss rates,
uncertain nuclear reaction rates and the treatment of convection and
mixing. Testing all of these uncertainties is outside the scope of this
study, and indeed would require an extremely large amount of work (it is
somewhat more tractable using synthetic stellar models, see eg.
\citealt{Marigo2001a}).  

Determining some bounds to AGB model results has however been attempted by
various studies. For instance our group explored the effects of varying
mass-loss rates and (some) nuclear reaction rates in low and intermediate
mass stars of low metallicity (\citealt{Fenner2004}).  A more detailed
study into the mass-loss dependence of AGB yields was recently reported by
\cite{Stancliffe2007a}, and \cite{Herwig2006} investigated the effect of
reaction rate uncertainties on TDU efficiency and the resultant AGB yields.
A further source of uncertainty arises because detailed AGB models are
currently lacking opacity tables that take into account the C and N
enrichment that occurs as a result of 3DU and HBB.  In particular low
temperature opacity tables variable in H, He, C and N (at least) are
needed. By approximating the opacity of a few key molecules in synthetic
stellar models, \cite{Marigo2002} showed how important including this
opacity source is. She found, amongst other things, that it affects the
effective temperature and lifetime of C-rich models. We note that
\cite{Lederer2008a} have just completed detailed calculations to produce
tables for use in AGB stellar models. Some preliminary results have
recently been used in a study by \cite{Cristallo2007}. We are currently
updating our structure code to include these opacity tables.

In the case of the Z=0 and EMP models considered here another source of
uncertainty arises -- that of handling the violent Dual Flash events.
\cite{Schlattl2001} investigated the dependence of the H-ingestion flashes
on physical inputs, finding that the occurrence of DFs is dependent on, for
example, the initial He abundance and the inclusion of atomic diffusion.
They did however conclude that DFs are a robust prediction of 1D stellar
models. They also noted that the degree and type of pollution arising from
their models is very similar to previous studies. It is reassuring that all
1D DCF simulations appear to give very similar results, despite the various
numerical methods employed between codes. However, seeing as most of the
models overestimate the amount of C and N production by orders of magnitude
(see \S\ref{compareobs}), this general agreement may just reflect a
consistent inadequacy of the 1D codes in handling the DCF events. For this
reason we believe that multidimensional fluid dynamics calculations are
really needed to model these violent episodes. This area of research is in
its infancy, however a couple of groups (eg. our group and
\citealt{Woodward2008}) have recently attempted preliminary simulations. As
the fluid dynamics calculations are limited to very short timescales -- a
simulation of a few hours of star-time takes $\sim1$ month on 32 processors
(private communication, Mocak 2008) -- the results from this work will
still need to be parameterised for use in 1D stellar codes. The fluid
dynamics simulations will eventually give important information on the
mixing velocities and the behaviour of convective boundaries in this
extreme environment, hopefully shedding light on the reason for the
overproduction of C and N in the 1D models.
%
%
\section{Summary}
We have modified an existing stellar code and used this to calculate the
evolution and nucleosynthesis of a grid of low- and intermediate-mass EMP
and Z=0 models, from the ZAMS to the end of the TPAGB. In this paper, the
first of a series to analyse the large amount of data produced, we have
presented the yields for the 20 stars. We also briefly analysed the C and N
yields in the context of the Galactic Halo CEMP population.

We find that the yields are universally C-rich, in qualitative agreement
with the CEMP observations. However, as previous studies have found, our
yields do not fit quantitatively -- most of the C yields are too high to
explain the most metal-poor objects ([Fe/H] $\lesssim -4.0$). The
disagreement is at the $\sim 1$ to 2 dex level, as also found previous
studies. Dilution of the yields through binary mass transfer or better
numerical modelling may resolve this discrepancy.  We also note that the
very small observational data sets at these extreme metallicities make
comparisons uncertain.

At higher metallicities ([Fe/H] $\sim -3.0$), where the observational
data set is much larger, we find that the C yields of our [Fe/H] = $-3.0$
models are compatible with the most C-rich CEMPs. Importantly it is the
models that have their yields dominated by the Dual Shell Flash pollution
(our `Group 2' yields) that best match the observations, based on the two
constraints considered here -- [C/Fe] and [C/N]. The models that have their
yields dominated by TDU and HBB produce comparable amounts of C but have
too much N (generated from the HBB), which pushes [C/N] to much lower
values than seen in the observations.  Moreover, if the lower mass stars
were to experience TDU (and no DFs), the pollution expected would be
(mainly) C-rich, whereas the CEMPs are enriched in N as well as C. Thus TDU
-- with or without HBB -- does not seem to be a viable solution for the
CEMP abundance patterns. We stress that the Dual Shell Flash events only
occur at low metallicities. We note that our result is in contrast with the
findings of \cite{Iwamoto2004} who found their [Fe/H] $=-2.7$ models to
produce $\sim 1$ to 2 dex too much C relative to the observations.

We also find that the models predict [C/Fe] to increase at lower and lower
metallicities. Furthermore, the proportion of CEMP stars should also
continue to increase at lower metallicities, based on the results that some
of the low mass EMP models already have polluted surfaces by the HB phase
(which has a long lifetime compared to the AGB), and that there are more
C-producing evolutionary episodes at these metallicities.

Finally we note that all these calculations contain many uncertainties.
These include the unknown mass-loss rates, uncertain nuclear reaction
rates, the treatment of convection and opacities. In the case of the Dual
Flash events we believe this warrants multidimensional fluid dynamics
calculations, which a number of groups have started working on.

The models and yields from this study will be described in more detail in
future papers in this series.
%
%
%
%
%
\begin{acknowledgements}
  This study utilised the Australian Partnership for Advanced Computing
  (APAC) supercomputer, under Project Code \textit{g61}.  SWC thanks the
  original authors and maintainers of the stellar codes that have been used
  in this work -- Peter Wood, Rob Cannon, John Lattanzio, Maria Lugaro,
  Amanda Karakas, Cheryl Frost, Don Faulkner and Bob Gingold. SWC was
  supported by a Monash University Research Graduate School PhD scholarship
  for 3.5 years.
\end{acknowledgements}
\bibliographystyle{aa} 
\bibliography{9597} 
%
%
\end{document}